\documentclass[]{spie}  

\usepackage{amsmath,amsfonts,amssymb}
\usepackage{graphicx}
\usepackage[colorlinks=true, allcolors=blue]{hyperref}

\title{The PolarKID project: polarization measurements with KIDs for the next generation of CMB telescopes}

\author[a*]{Sofia Savorgnano}
\author[a]{Julien Bounmy}
\author[a]{Olivier Bourrion}
\author[b]{Martino Calvo}
\author[a]{Andrea Catalano}
\author[a]{Olivier Choulet}
\author[b]{Gregory Garde}
\author[b]{Anne Gerardin}
\author[a]{Mile Kusulja}
\author[a]{Juan Francisco Macías-Pérez}
\author[b]{Alessandro Monfardini}
\author[a]{Damien Tourres}
\author[a]{Francis Vezzu}
\affil[a]{LPSC, CNRS-IN2P3/Université Grenoble Alpes, 53 Av. des Martyrs, Grenoble 38000 (FR)}
\affil[b]{Institut Néel, CNRS, 25 Av. des Martyrs, Grenoble 38000 (FR)}

\authorinfo{*Corresponding author: savorgnano@lpsc.in2p3.fr}

\begin{document} 
\maketitle

\begin{abstract}
The goal of the PolarKID project is testing a new method for the measurement of polarized sources, in order to identify all the possible instrumental systematic effects that could impact the detection of CMB B-modes of polarization. It employs the KISS (KIDs Interferometer Spectrum Survey) instrument coupled to a sky simulator and to sources such as point-like black bodies (simulating planets), a dipole (extended source) and a polarizer. We use filled-arrays Lumped Element Kinetic Inductance Detectors (LEKIDs) since they have multiple advantages when observing both in a photometry and in a polarimetry configuration. 
\end{abstract}


\section{INTRODUCTION}
\label{sec:intro}  

The ambitious goal of detecting CMB B-modes of polarization requires much preliminary work in order to characterize and isolate eventual systematic effects that could impact their detection. This is why the project PolarKID has been launched and is currently being financed by the French Space Agency (CNES), between 2023 and 2025. The main goal of this project is testing in laboratory the ability of filled array KIDs to control systematic effects in polarimetry. To do so, we use the KISS (KIDs Interferometer Spectrum Survey) instrument (Ref.~\citenum{Fasano2022}), previously installed at the 2.25 m QUIJOTE telescope in Tenerife and now in the cryogenic laboratory of LPSC in Grenoble. Optically coupled to KISS, we use a sky simulator: a black body that emits similarly to the CMB and the atmosphere combined. In front of it, we place interchangeable sources such as point-like black bodies (simulating a photometric calibration source), a dipole (fully polarised point source) or an extended polarised source. The instrument can be used in different configurations: as a photometer, as a polarimeter or as an FTS (Fourier Transform Spectrometer). All the components of the experimental setup are studied and tested: from the fabrication of efficient cryogenic detectors, the Lumped Element Kinetic Inductance Detectors (LEKIDs), to the characterization of optical elements such as the silicon lenses. The LEKIDs detectors used in the so-called filled array configuration have multiple advantages: a very high quantum efficiency in a 30\% mm-band, a large filling factor and a relatively easy fabrication process (Ref.~\citenum{Catalano2020}). Moreover, they show promising results from the NIKA2 instrument when observing polarization: a sensitivity of $\sim 20 ~ mJ\sqrt{s}$ and a polarisation leakage $< $ 1\% (Ref.~\citenum{Perotto2020}). For all these reasons, we are developing cutting-edge KIDs technology that will play a key role in the development of the new telescopes dedicated to the detection of the polarization B-modes, such as the recently proposed French SAT for the Simons Observatory.  

\newpage
\section{EXPERIMENTAL SETUP}
\label{sec:setup}

\begin{figure}[h]
    \centering
    \includegraphics[width=0.8\textwidth]{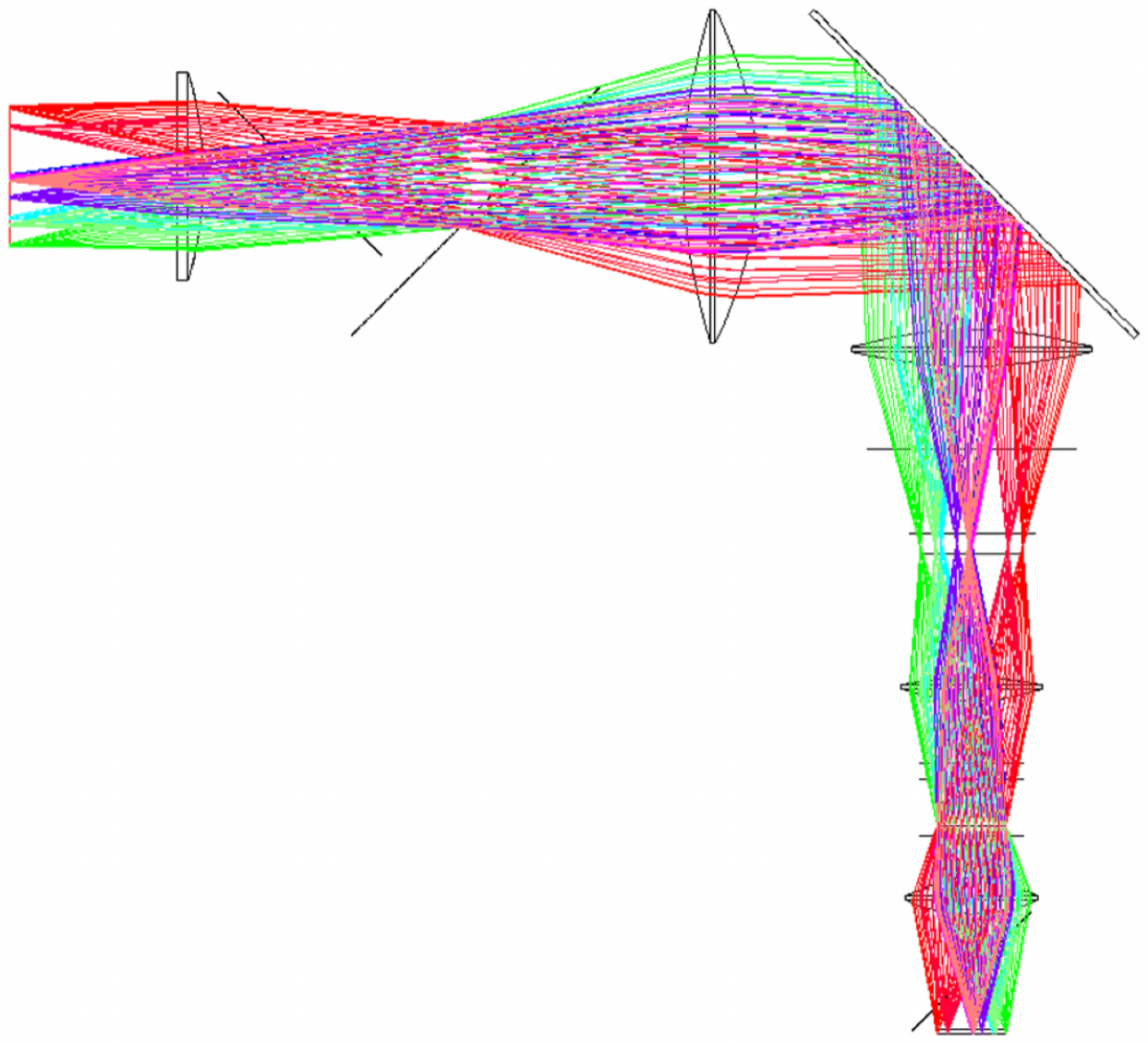}
    \caption{Zemax drawing of the experimental setup in a photometry configuration.}
    \label{fig:KISS}
\end{figure}

The instrumental setup is composed of the former KISS instrument coupled to a sky simulator, which reproduces the combined signal of the sky and atmosphere. In particular, the KISS cryostat is hosting multiple cold stages, the coldest of which is reaching a temperature of 140 mK. On the other hand, the sky simulator is a black body cooled to a nominal temperature of 21 K, which emits at $\sim$ 29 K due to the incremental contribution of optical elements such as lenses. Fig.~\ref{fig:KISS} shows a Zemax drawing of the experimental setup in the photometry configuration, with optical elements such as focusing lenses and the reflecting mirror. The signal from the sky simulator is represented by rays incoming from the left side of the drawing and focused on the focal plane, on the bottom of the drawing.

\begin{figure}
    \centering
    \begin{minipage}{0.45\textwidth}
        \centering
        \includegraphics[height=1\textwidth]{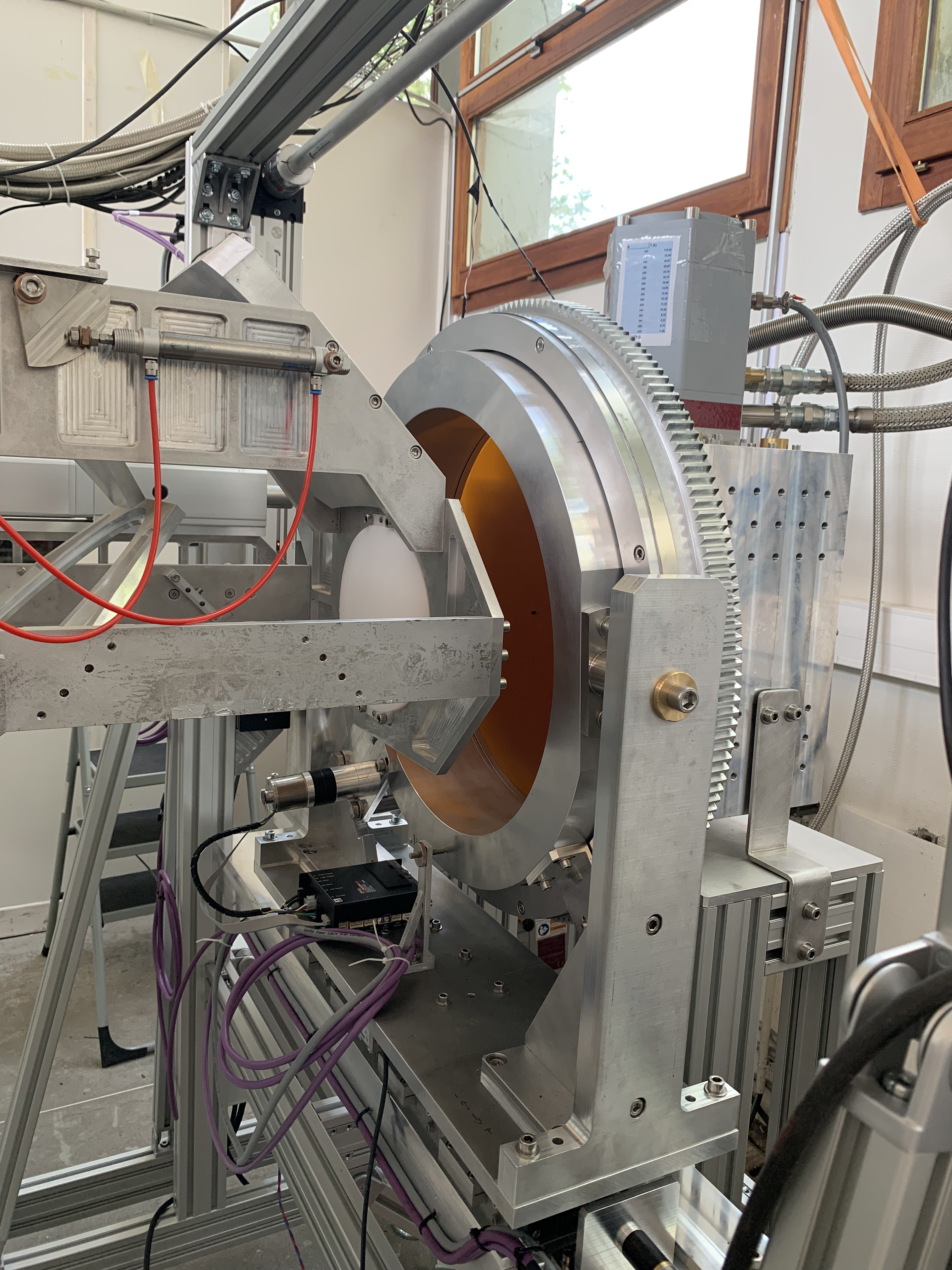}
        \caption{Picture of the sky simulator structure.}
        \label{fig:sky_sim}
    \end{minipage}\hfill
    \begin{minipage}{0.45\textwidth}
        \centering
        \includegraphics[height=1\textwidth]{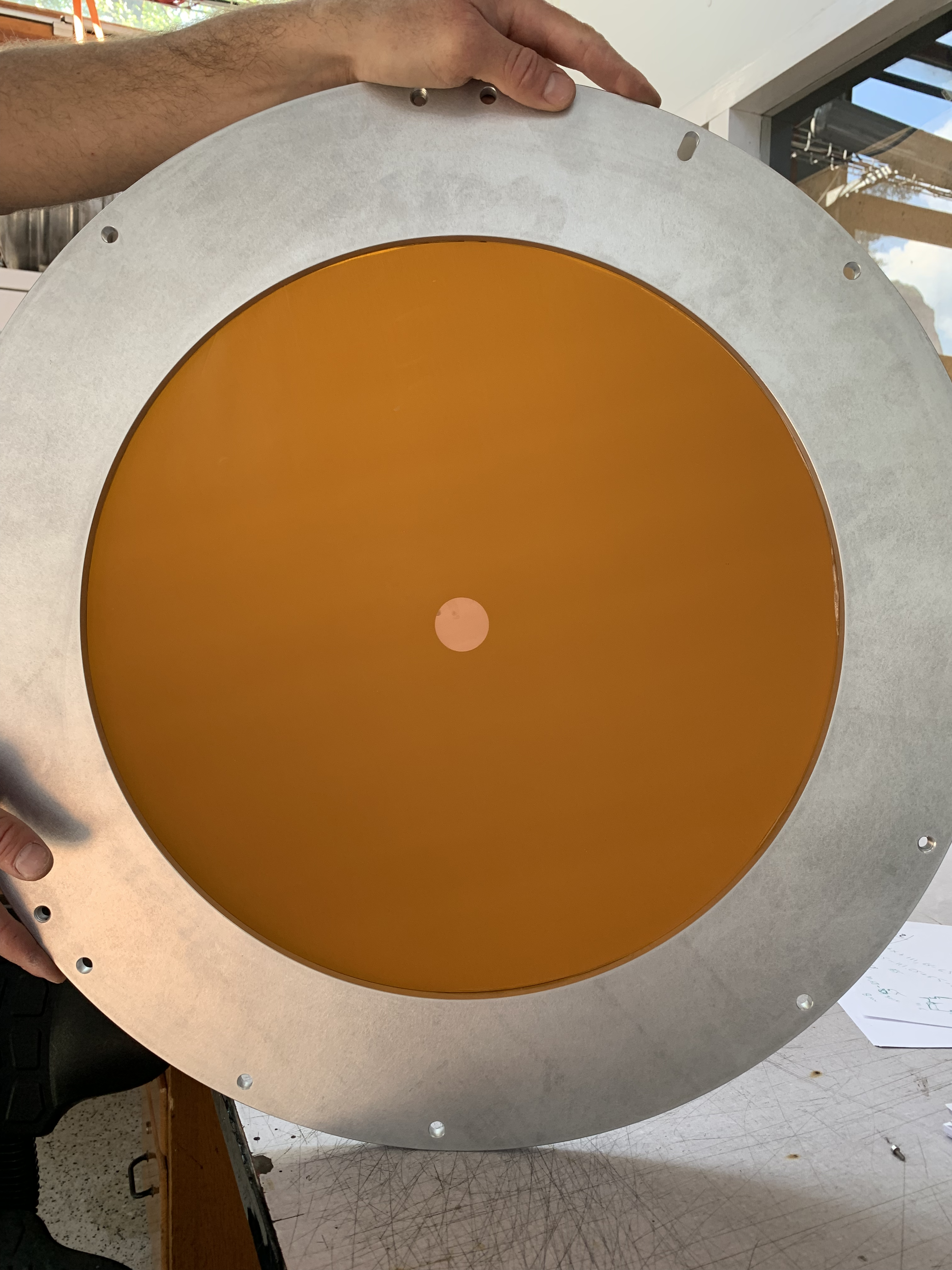}
        \caption{Picture of the disk with attached film hosting the polarizer.}
        \label{fig:polarizer}
    \end{minipage}
\end{figure}

In front of the sky simulator, we place a metallic ring filled with a kapton film (Fig.~\ref{fig:sky_sim}). Depending on the type of measurement we wish to perform, a specific source is installed on the structure, so overall we can interchange 3 different disks. We dispose of a photometric source, meant to simulate a photometric point-like source, which is essentially a copper disk with a black layer on it, of 3 mm in diameter, glued at the center of the kapton film. A second source we want to test is an elongated wire (3 mm long), which acts as a 3/2$\lambda$ dipole producing a fully polarized emission at 150 GHz. Finally, we also want to simulate an extended diffuse source, which we do with a 3 cm linear polarizer, directly imprinted onto the kapton film (Fig.~\ref{fig:polarizer}).

The structure supporting the source can be moved in both (X,Y) coordinates and the disk can turn, either by a selected angle or by continuous rotation, with a precision of 3 arc-minutes. The user can choose to move the source by manual mode or by performing a scan, with the desired dimension, speed and rotation frequency. All the scanning parameters can be selected through a dedicated software. The advantage of this setup is the accuracy in reproducing the conditions we have when observing the sky, except for the fact that in this case, we have a fix ``telescope" and a scanning source. 

\section{KID ARRAY CHARACTERIZATION}
\label{sec:kids}

\begin{figure}
    \centering
    \includegraphics[width=0.5\textwidth]{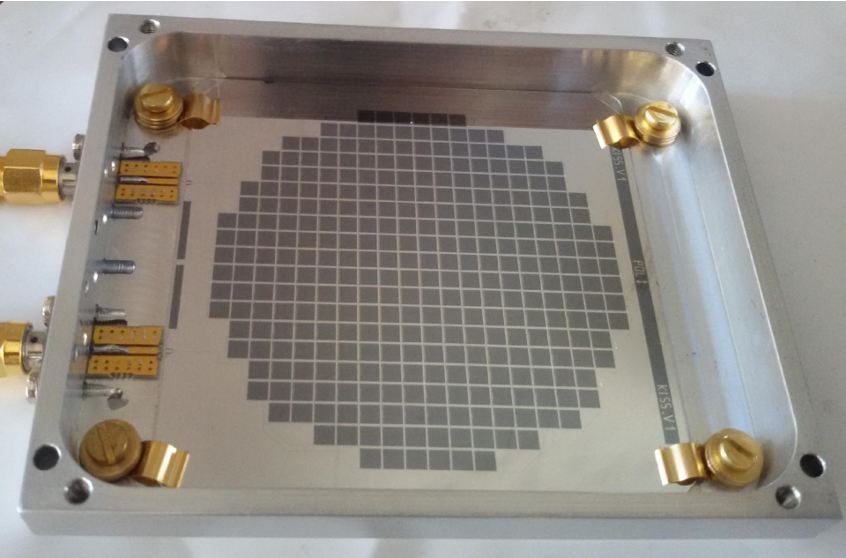}
    \caption{Picture of the KIDs array used for lab measurements.}
    \label{fig:array}
\end{figure}

\begin{figure}
    \centering
    \includegraphics[width=0.9\textwidth]{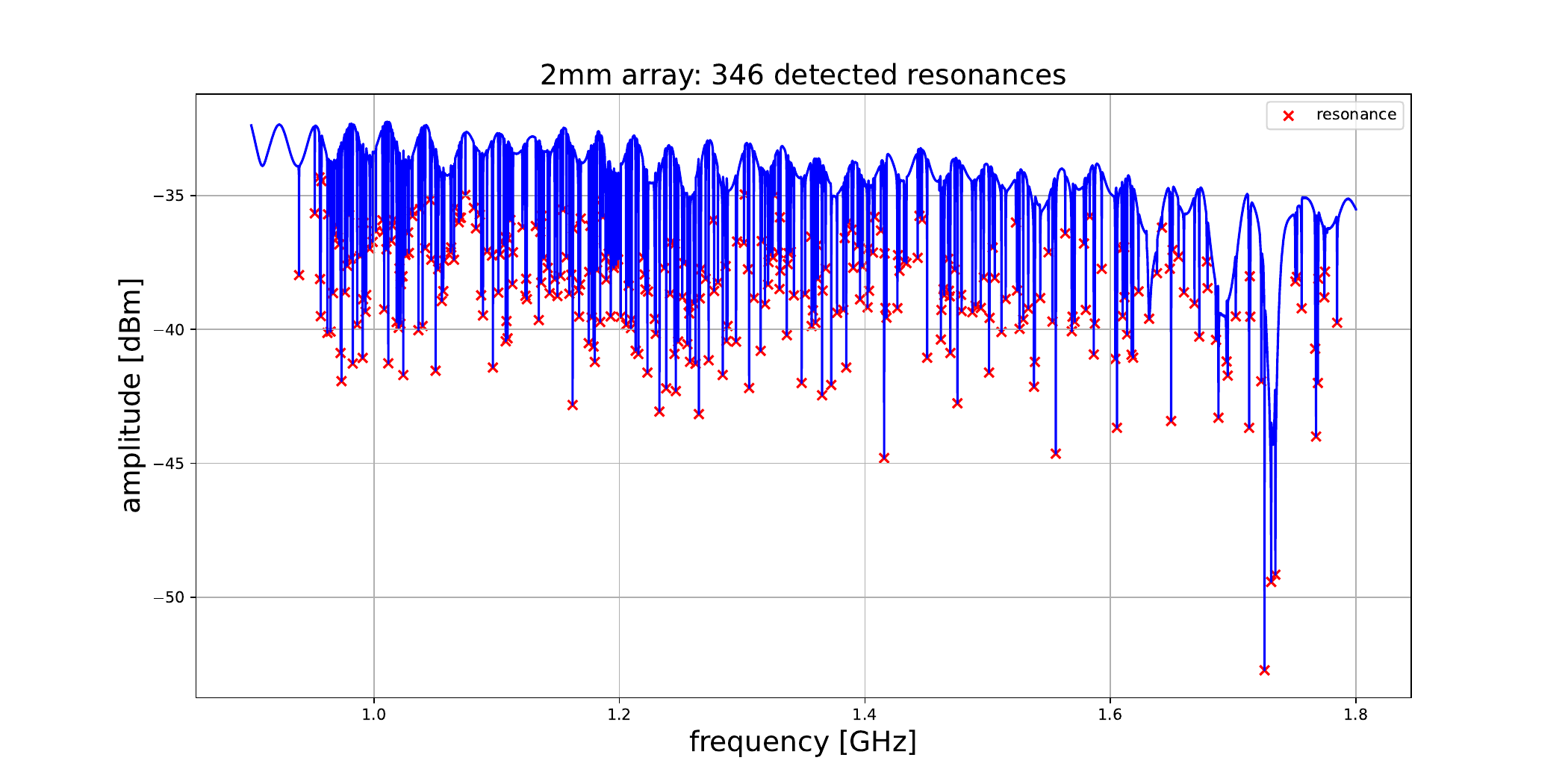}
    \caption{Feedline of the KIDs array acquired through a VNA scan. The red crosses identify the resonance frequencies for each pixel.}
    \label{fig:feeline}
\end{figure}

One of the main assets in this measurement campaign is the use of Kinetic Inductance Detectors, which are increasingly used for ground-based instruments devoted to the observation of mm-wavelength radiation. The arrays used for these measurements are realized on a 300 $\mu$m thick silicon wafer and deposited with 25 nm of aluminium as superconductor substrate (Fig.~\ref{fig:array}). The two arrays are composed of 418 pixels, each with a specific resonance frequency ranging from 900 MHz to 1.2 GHz (Fig.~\ref{fig:feeline}). The wavelength range of sensitivity is 240 - 260 GHz, peaked at 250 GHz. The arrays are placed at the coldest cryogenic stage, oriented perpendicularly with respect to each other and the light beam illuminating them is split by a linear polarizer oriented at 45º with respect to the incoming line of sight. 

Before being used for measurements, the arrays have been tested in order to quantify some fundamental parameters and therefore assess their quality and performance. By fitting the resonances shown in Fig.~\ref{fig:feeline}, we can derive information about the quality of the array. In Tab.~\ref{tab:perf_kids} the main characteristic performances for the arrays used in the measurements campaign are listed. 

\begin{table}[h]
    \centering
    \begin{tabular}{|c|c|c|}
    \hline
     & NICA V10.1 & NICA V10.2\\
    \hline
    Silicon wafer & 321 $\mu$m  & 321 $\mu$m\\
    Central frequency & 150 GHz & 150 GHz\\
    Resonances & 87\% & 83\%\\
   $Q_{tot}$ & 16555 $\pm$ 4232 & 18340 $\pm$ 3209\\
    Responsivity & (641 $\pm$ 80) Hz/K & (720 $\pm$ 90) Hz/K\\
    Noise & (11 $\pm$ 2) $Hz/\sqrt{Hz}$ & (12 $\pm$ 2) $Hz/\sqrt{Hz}$\\
    \hline
    \end{tabular}
\caption{Performances of the KIDs array used for laboratory tests.}
\label{tab:perf_kids}
\end{table}

\section{MEASUREMENT METHODS AND GOALS}
\label{sec:meas}

Two separated measurement sessions have been performed, with different sources used and different goals. In the first place, we conducted photometric measurements, so using the circular black body simulating a planet as a source. Then, we replaced the photometric source with the extended polarized source. Even though the measuring technique is quite similar, the results and goals are very different in the two cases.

\subsection{Photometry measurements}

\begin{figure}
    \centering
    \includegraphics[width=0.8\textwidth]{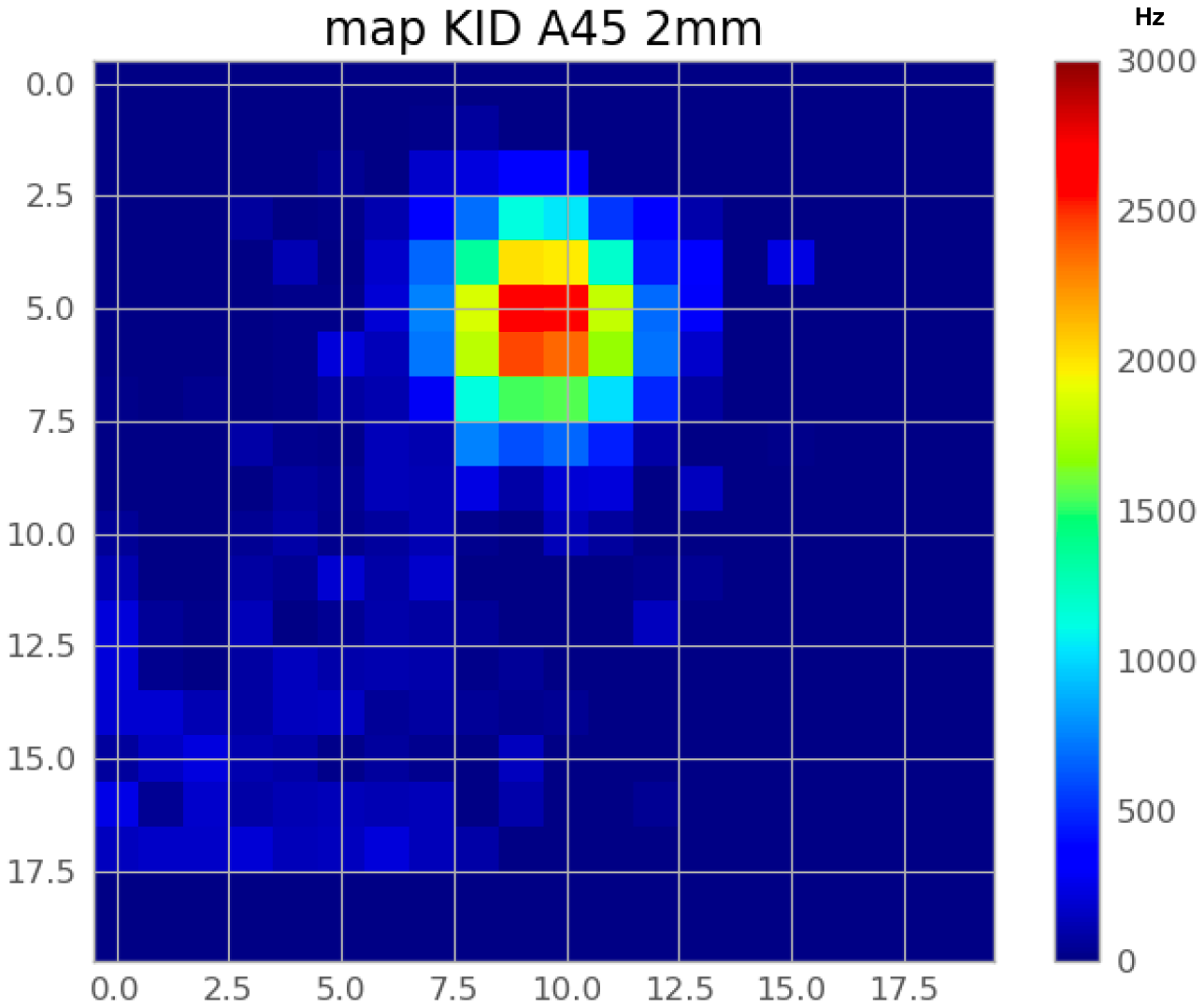}
    \caption{Projected map of the point-like source (planet) as seen by one pixel of the reflection array. The x and y axes are simply representing the map's pixels, while the colorbar shows the intensity of the signal expressed in Hz.}
    \label{fig:map_planet}
\end{figure}

The first step is performing photometric measurements of a point-like source, which is meant to simulate a photometric calibration source, such as a planet. This source, placed at the ambient temperature of 300 K, is diluted by the beam (which is 12 mm) by a factor 6, therefore the temperature variation produced with respect to the optical background is 50 K. Such a signal should produce a shift in the resonance frequency of detectors of about 3 kHz. This frequency shift is the direct quantity that is registered by detectors when a hotter source is illuminating them. In other words, the amplitude of the shift of a detector's resonance frequency is proportional to the intensity of the signal that this detector is receiving. Therefore, from the frequency shift, we can derive time ordered data that show the amplitude of the shift in Hz as the source is scanning the focal plane. Finally, by projecting this timeline data on a plane, we can create an actual map of the source and gather information on its amplitude, position as seen by every single pixel and beam of pixels. By fitting this projected map, as the one shown in Fig.~\ref{fig:map_planet}, we obtain an estimation of the center in (X,Y) coordinates, the FWHM or beam of the pixel, as well as other parameters like the eccentricity and the amplitude. 

By repeating the fitting procedure for all pixels of the two arrays, we can reconstruct the position of each pixel on the focal plane and so derive information on the geometry of the array, on the image plane. Moreover, since the fit of the map is also providing the FWHM, we can derive the beam for each pixel of both arrays. The beam averaged on all pixels of the arrays is estimated to be $12 \pm 2$ mm. 

\subsection{Polarimetry measurements}
The second, and more crucial, step of the PolarKID project is related to the measurement of polarization and the characterization of the systematic effects that limit the precision within which we can determine the polarization angle of the emitting source. The measurement method is similar to the one adopted in photometry, meaning that we perform a scan of the source spanning the whole focal plane. However, this time we make use of the extended polarised source and, while scanning, the source is rotating at a constant rotational frequency. This continuous rotation is adding a modulation to the signal that we expect to observe. 
Before performing the actual measurements, it is essential to model the system in order to understand which variables are at play and which kind of signal we should expect. The main variables to take into account are the scanning speed, the average beam (value derived from photometry: 12 $\pm$ 2 mm), polarizer's diameter (30 mm), position of the pixel in the focal plane (derived from geometry). With these parameters, we can create a Gaussian window function, with an effective beam given by $FWHM = \sqrt{fw_{pix}^2 + d^2}$, where $fw_{pix}$ is the beam of a pixel and d is the polarizer's diameter. Then, we can multiply this window function by the polarized signal calculated through Mueller matrices. The polarized signal seen by detectors will be given by the application of three matrices to one initial Stokes vector of unpolarized light (S = [1,0,0]). The first matrix is a Mueller matrix for an ideal linear polarizer (Eq.~\ref{eq:mueller}), where the angle $\alpha$ is the orientation of the wires with respect to the incoming light direction. The second matrix accounts for the rotation provided by a mirror, oriented in principle at 45º, reflecting the signal towards the cryostat's aperture. The third matrix is another Mueller matrix, which takes into account the polarizer placed at the 100 mK stage and responsible for splitting the signal on the two arrays. 

\begin{equation}
    M = \frac{1}{2} \begin{bmatrix}
        1 & \cos 2\alpha & \sin 2\alpha  \\
        \cos 2\alpha & \cos^2 2\alpha & \cos 2\alpha \sin 2\alpha  \\
        \sin 2\alpha & \cos 2\alpha \sin 2\alpha & \sin^2 2\alpha
    \end{bmatrix} 
    \label{eq:mueller}
\end{equation}

By applying these matrices to the initial Stokes vector, we obtain Eq.\ref{eq:I_fin}, where $\beta$ is the splitting polarizer orientation angle, $\theta$ is the mirror's inclination angle and $\alpha$ is the polarized source's angle. 

\begin{equation}
    I = 1 + cos 2\beta [sin (2\theta + 2\alpha)] + sin 2\beta [cos (2\theta + 2\alpha)]
    \label{eq:I_fin}
\end{equation}

\begin{figure}
    \centering
    \includegraphics[width=0.8\textwidth]{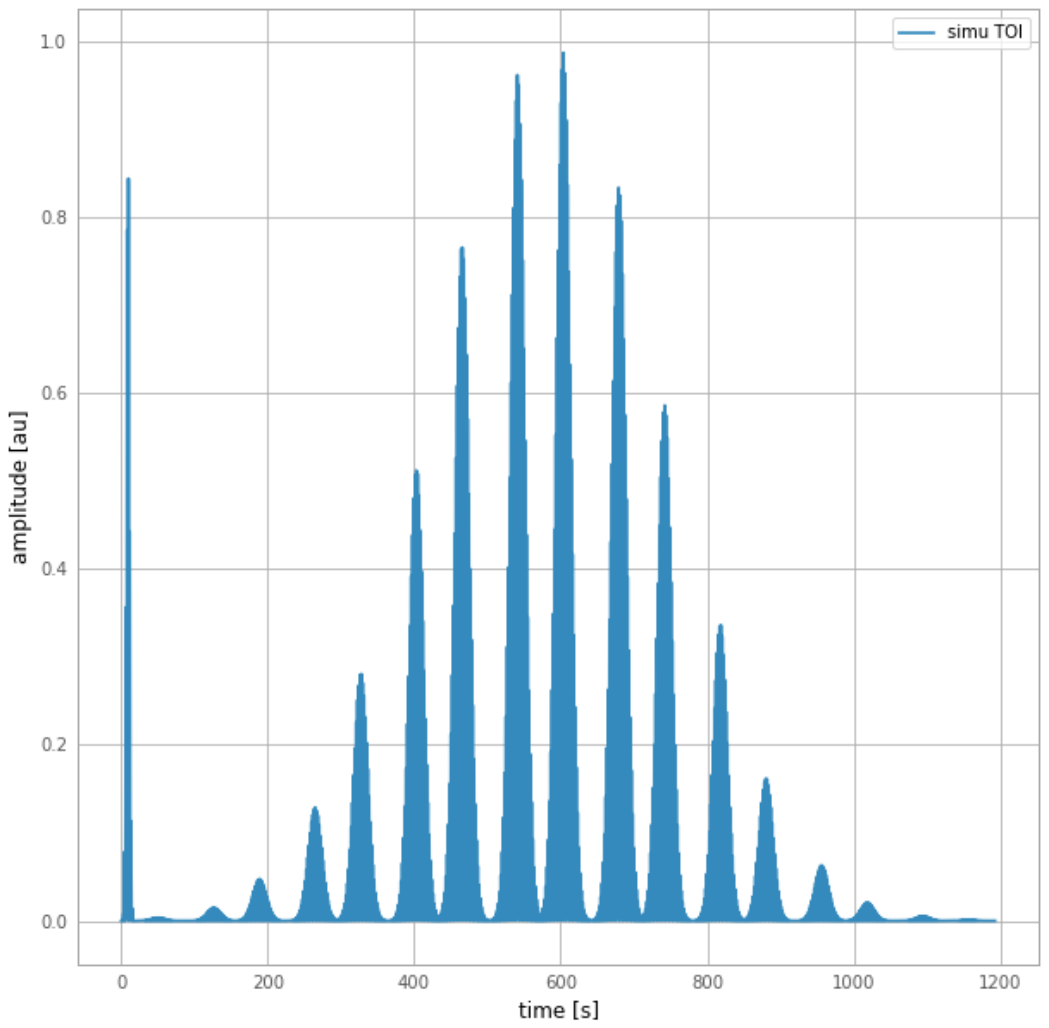}
    \caption{Simulated timeline for one selected pixel of the transmission array, in the case of a continuous rotation of the source at 103º/s.}
    \label{fig:simu_toi}
\end{figure}

The final polarized signal that we expect to see in one pixel's timeline is shown in Fig.~\ref{fig:simu_toi}.

The following step will be a comparison between the simulated timeline and real data. In principle, by fitting real polarized data with this simulated model, we can derive an estimation of the orientation angle of the polarized source, which would give us the limiting precision due to systematic effects. Then, with this modulated timeline, we can separate the intensity and polarization components, in order to project I, Q and U polarization maps. This is a standard procedure for astrophysical maps of polarized sources, so we would be able to reproduce this analysis for an in-lab, more controlled system. 

\begin{figure}
    \centering
    \includegraphics[width=0.7\textwidth]{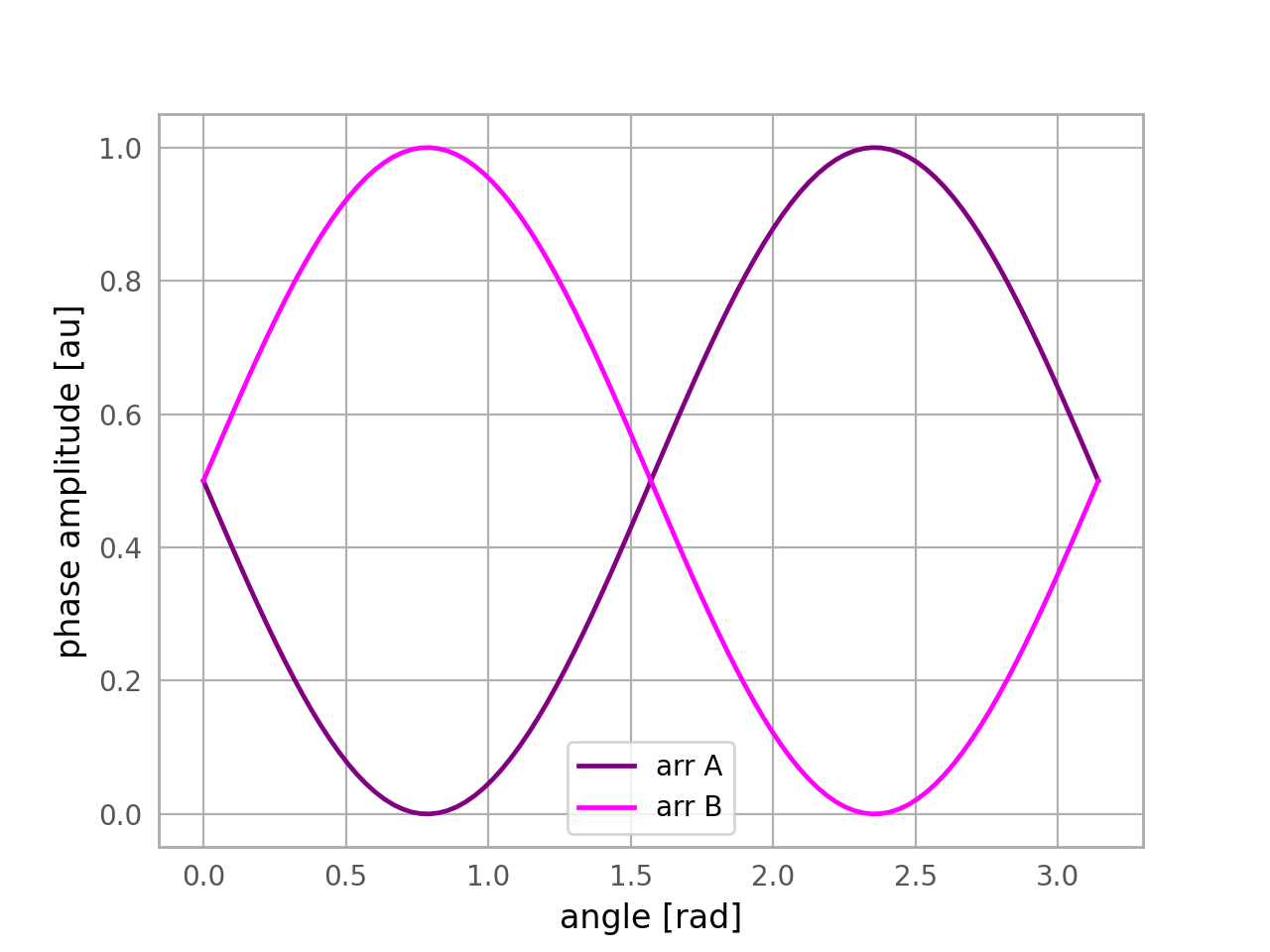}
    \caption{Amplitude of the signal for the two arrays, in reflection (A) and in transmission (B), for orientation angles of the source in the range from 0º to 180º.}
    \label{fig:step_curve}
\end{figure}

Another method of measuring polarization with this kind of experimental setup is given by a step-motor technique: in this case, the polarizer will not continuously rotate, but it will change its position by a finite angle for each scan. Therefore, we will not have any modulation, but we can reconstruct the amplitude of the signal for each different scan and fit the resulting curve with the model (Eq.\ref{eq:I_fin}), which is common for both cases. A simulated curve of the resulting fit is shown in Fig.~\ref{fig:step_curve}. 

\section{CONCLUSION AND PERSPECTIVES}
\label{sec:conclusion}

Through this measurement techniques, we can derive direct information on the polarization angle of the source and the associated error. Therefore, we can obtain a limit on the precision within which the polarization angle is known. This project's main goal is proving that our instrumental setup is capable of detecting polarization with the precision required for upcoming CMB experiments devoted to B-modes. In particular, we make use of filled array detectors, illuminated by a signal split by a 45º polarizer. This configuration is rather different with respect to horn-coupled detectors, but in principle it is equally fit to observe CMB polarization. Moreover, it presents multiple advantages, among which is the ability of collecting all incident photons as we are filling the whole focal plane with sensitive detectors. 
Gathering information about the precision within which we can know a source's polarization angle is crucial for any experiment aiming at observing cosmological B-modes, since the calibration of such instruments is a delicate yet essential step. In fact, the precision associated to the polarization angle is directly related to the estimated value of the tensor-to-scalar ratio $r$, the variable describing the amplitude of primordial B-modes. Currently, the value predicted by the theory is $r = 0.004$, which means that any measurement can only tolerate a calibration error $<$ 0.1º. The final goal and last missing step of the PolarKID project is precisely proving that this measurement method assures this precision on the polarization angle, making the KIDs filled array configuration suitable for future experiments willing to detect cosmological polarization.


\acknowledgments 
 
This work has been partially supported by the LabEx FOCUS ANR-11-LABX-0013. The authors acknowledge the contribution of CNES to the R\&D project ``PolarKID: mesure de polarisation du CMB par détecteurs LEKIDs". 

\bibliography{report} 

\begin{thebibliography}{1}

\bibitem{Fasano2022}
{Fasano}, A., {Catalano}, A., {Mac{\'\i}as-P{\'e}rez}, J.~F., {Aguiar}, M., {Beelen}, A., {Benoit}, A., {Bideaud}, A., {Bounmy}, J., {Bourrion}, O., {Bres}, G., {Calvo}, M., {Castro-Almaz{\'a}n}, J.~A., {de Bernardis}, P., {De Petris}, M., {de Taoro}, A.~P., {Fern{\'a}ndez-Torreiro}, M., {Garde}, G., {G{\'e}nova-Santos}, R., {Gomez}, A., {G{\'o}mez-Renasco}, M.~F., {Goupy}, J., {Hoarau}, C., {Hoyland}, R., {Lagache}, G., {Marpaud}, J., {Marton}, M., {Monfardini}, A., {Peel}, M.~W., {Pisano}, G., {Ponthieu}, N., {Rebolo}, R., {Roudier}, S., {Rubi{\~n}o-Mart{\'\i}n}, J.~A., {Tourres}, D., {Tucker}, C., and {Vescovi}, C., ``{Observations with KIDs Interferometer Spectrum Survey (KISS)},'' {\em European Physical Journal Web of Conferences} {\bf 257},  00017 (July 2022).

\bibitem{Catalano2020}
{Catalano}, A., {Bideaud}, A., {Bourrion}, O., {Calvo}, M., {Fasano}, A., {Goupy}, J., {Levy-Bertrand}, F., {Mac{\'\i}as-P{\'e}rez}, J.~F., {Ponthieu}, N., {Tang}, Q.~Y., and {Monfardini}, A., ``{Sensitivity of LEKID for space applications between 80 GHz and 600 GHz},'' {\em AAP}~{\bf 641},  A179 (Sept. 2020).

\bibitem{Perotto2020}
{Perotto}, L., {Ponthieu}, N., {Mac{\'\i}as-P{\'e}rez}, J.~F., {Adam}, R., {Ade}, P., {Andr{\'e}}, P., {Andrianasolo}, A., {Aussel}, H., {Beelen}, A., {Beno{\^\i}t}, A., {Berta}, S., {Bideaud}, A., {Bourrion}, O., {Calvo}, M., {Catalano}, A., {Comis}, B., {De Petris}, M., {D{\'e}sert}, F.~X., {Doyle}, S., {Driessen}, E.~F.~C., {Garc{\'\i}a}, P., {Gomez}, A., {Goupy}, J., {John}, D., {K{\'e}ruzor{\'e}}, F., {Kramer}, C., {Ladjelate}, B., {Lagache}, G., {Leclercq}, S., {Lestrade}, J.~F., {Maury}, A., {Mauskopf}, P., {Mayet}, F., {Monfardini}, A., {Navarro}, S., {Pe{\~n}alver}, J., {Pierfederici}, F., {Pisano}, G., {Rev{\'e}ret}, V., {Ritacco}, A., {Romero}, C., {Roussel}, H., {Ruppin}, F., {Schuster}, K., {Shu}, S., {Sievers}, A., {Tucker}, C., and {Zylka}, R., ``{Calibration and performance of the NIKA2 camera at the IRAM 30-m Telescope},'' {\em AAP}~{\bf 637},  A71 (May 2020).

\end{thebibliography}
\bibliographystyle{spiebib} 

\end{document}